\documentclass[twocolumn,fp]{jpsj3}
\usepackage{graphicx}
\usepackage{bm}
\usepackage{cite}
\usepackage{color}
\usepackage{amssymb}
\usepackage{amsmath}
\usepackage{revsymb}

\title{Phase Diagram of the Frustrated Square-Lattice Hubbard Model:\\ Variational Cluster Approach}
\author{Kazuma Misumi, Tatsuya Kaneko, and Yukinori Ohta}
\inst{Department of Physics, Chiba University, Chiba 263-8522, Japan}

\date{\today}

\abst{
The variational cluster approximation is used to study the frustrated Hubbard model 
at half filling defined on the two-dimensional square lattice with anisotropic 
next-nearest-neighbor hopping parameters.  We calculate the ground-state 
phase diagrams of the model in a wide parameter space for a variety of lattice 
geometries, including square, crossed-square, and triangular lattices.  
We examine the Mott metal-insulator transition and show that, in the Mott 
insulating phase, magnetic phases with N\'eel, collinear, and spiral orders appear in 
relevant parameter regions, and in an intermediate region between these phases, 
a nonmagnetic insulating phase caused by the quantum fluctuations in the geometrically 
frustrated spin degrees of freedom emerges.  
}

\begin{document}
\maketitle

\section{Introduction}

The effect of geometrical frustration in strongly correlated electron systems has 
been one of the major issues of condensed matter physics.  In particular, a spin-liquid 
state caused by the frustration has been interpreted as an exotic state of matter, 
where the magnetic long-range order is destroyed, yielding a quantum paramagnetic 
(or nonmagnetic) state at zero temperature \cite{balents} or even exotic mechanisms 
of high-temperature superconductivity \cite{lee}.  
The Hubbard, Heisenberg, and related models defined on two-dimensional square 
and triangular lattices with geometrical frustration have been studied in this respect 
to find novel quantum disordered states by a variety of theoretical methods.  

In the square-lattice cases, the $J_1$-$J_2$ Heisenberg model with the nearest-neighbor 
($J_1$) and next-nearest-neighbor ($J_2$) exchange interactions have been studied 
for more than two decades \cite{chandra,sachdev,mario,chubukov,oitmaa,zhitomirsky,
trumper1,singh,capriotti1,capriotti2,zhang,takano,sirker,mambrini,darradi,isaev,beach,
richter,reuther1,yu1,wang1,jiang2,mezzacapo,li1,wang2,hu,doretto,qi,gong,morita1}. 
At $J_2=0$, where the frustration is absent, the model is known to have the N\'eel-type 
antiferromagnetic long-range order.  With increasing $J_2$, the frustration increases, 
but at $J_2=J_1$, the model again has the ground state with the collinear antiferromagnetic 
long-range order.  The strongest frustration occurs around $J_2/J_1=0.5$, where 
nonmagnetic states such as a valence bond state 
\cite{sachdev,chubukov,zhitomirsky,singh,capriotti1,takano,sirker,mambrini,isaev,yu1,doretto} 
and a spin-liquid state \cite{capriotti2,jiang2,mezzacapo,li1,wang2,hu} have been suggested 
to appear, the region of which has recently been studied further in detail \cite{gong,morita1}.  
The $t_1$-$t_2$-$U$ Hubbard model with the nearest-neighbor ($t_1$) and 
next-nearest-neighbor ($t_2$) hopping parameters and the on-site repulsive interaction 
$U$ has also been studied, where it has been shown that the critical interaction strength 
$U_c$ of the metal-insulator transition increases monotonically with increasing 
$t_2/t_1$ \cite{mizusaki,nevidomskyy,yu2} and that the ground state has the N\'eel order 
at a small $t_2/t_1$ and a collinear order around $t_2=t_1$ \cite{mizusaki,yu2}.  
Then, the nonmagnetic insulating state appears between these ordered states 
\cite{nevidomskyy,yamaki}.  

In the triangular-lattice cases, the anisotropic $J$-$J'$ triangular Heisenberg model has 
been studied.  In the isotropic case ($J=J'$), the 120$^\circ$ spiral ordered phase is known 
to be stable \cite{sindzingre}.  In the anisotropic case, the N\'eel order is realized when 
$J'/J$ is small and the spiral order is realized around $J'/J=1$ 
\cite{merino1,weihong,trumper2,yunoki,weng,starykh,bishop,reuther2,weichselbaum,hauke1,hauke2,merino2}, 
and between these phases, a dimer ordered phase \cite{weihong} or a spin-liquid phase 
\cite{hauke1,hauke2} has been predicted to appear.  
The anisotropic $t$-$t'$-$U$ triangular Hubbard model has also been studied 
\cite{morita,sahebsara,watanabe,tocchio1,tocchio2,yamada,laubach}, where it has been shown 
that $U_c$ increases with increasing $t'/t$: at a small $t'/t$ a metal-insulator transition 
occurs from the metallic phase to the N\'eel ordered phase, whereas at $t'/t\simeq 1$ a 
nonmagnetic insulating phase appears between the metallic and spiral ordered phases 
\cite{yamada,laubach}.  
Recently, the magnetic orders in the triangular-lattice Heisenberg model with $J_1$ and 
$J_2$ have also been studied, where a nonmagnetic insulating phase is shown to appear 
between the spiral and collinear phases \cite{mishmash,kaneko,li2,zhu,hu2}.  

In this paper, motivated by the above developments in the field, we study the frustrated 
square-lattice Hubbard model at half filling with the isotropic nearest-neighbor and 
anisotropic next-nearest-neighbor hopping parameters and clarify the metal-insulator transition, 
the appearance of possible magnetic orderings, and the emergence of a nonmagnetic insulating 
phase.  The search is made in a wide parameter space including square, crossed-square, and 
triangular lattices, as well as in weak to strong electron correlation regimes.  
We use the variational cluster approximation (VCA) based on self-energy functional 
theory (SFT) \cite{potthoff1,potthoff2,dahnken}, which enables us to take into account 
the quantum fluctuations of the system, so that we can study the effect of geometrical 
frustration on the spin degrees of freedom and determine the critical interaction 
strength for the spontaneous symmetry breaking of the model.  We examine the entire 
regime of the strength of electron correlations at zero temperature, of which little detail 
is known.  In particular, we compare our results in the strong correlation regime with those 
of the Heisenberg model, for which many studies have been carried out.  
We also compare our results with those in the weak correlation limit via the generalized 
magnetic susceptibility calculation and with those of the classical Heisenberg model 
calculation where the quantum spin fluctuations are absent.  

We will thereby show that magnetic phases with N\'eel, collinear, and spiral orders appear 
in relevant regions of the parameter space of our model and that a nonmagnetic insulating 
phase, caused by the quantum fluctuations in the frustrated spin degrees of freedom, 
emerges in a wide parameter region between the ordered phases obtained.  
The orders of the phase transitions will also be determined.  We will summarize our results 
as a ground-state phase diagram in a full two-dimensional parameter space.  This phase 
diagram will make the characterization of the nonmagnetic insulating phase more approachable, 
although this is beyond the scope of the present paper.  

\begin{figure}[thb]
\begin{center}
\includegraphics[width=0.65\columnwidth]{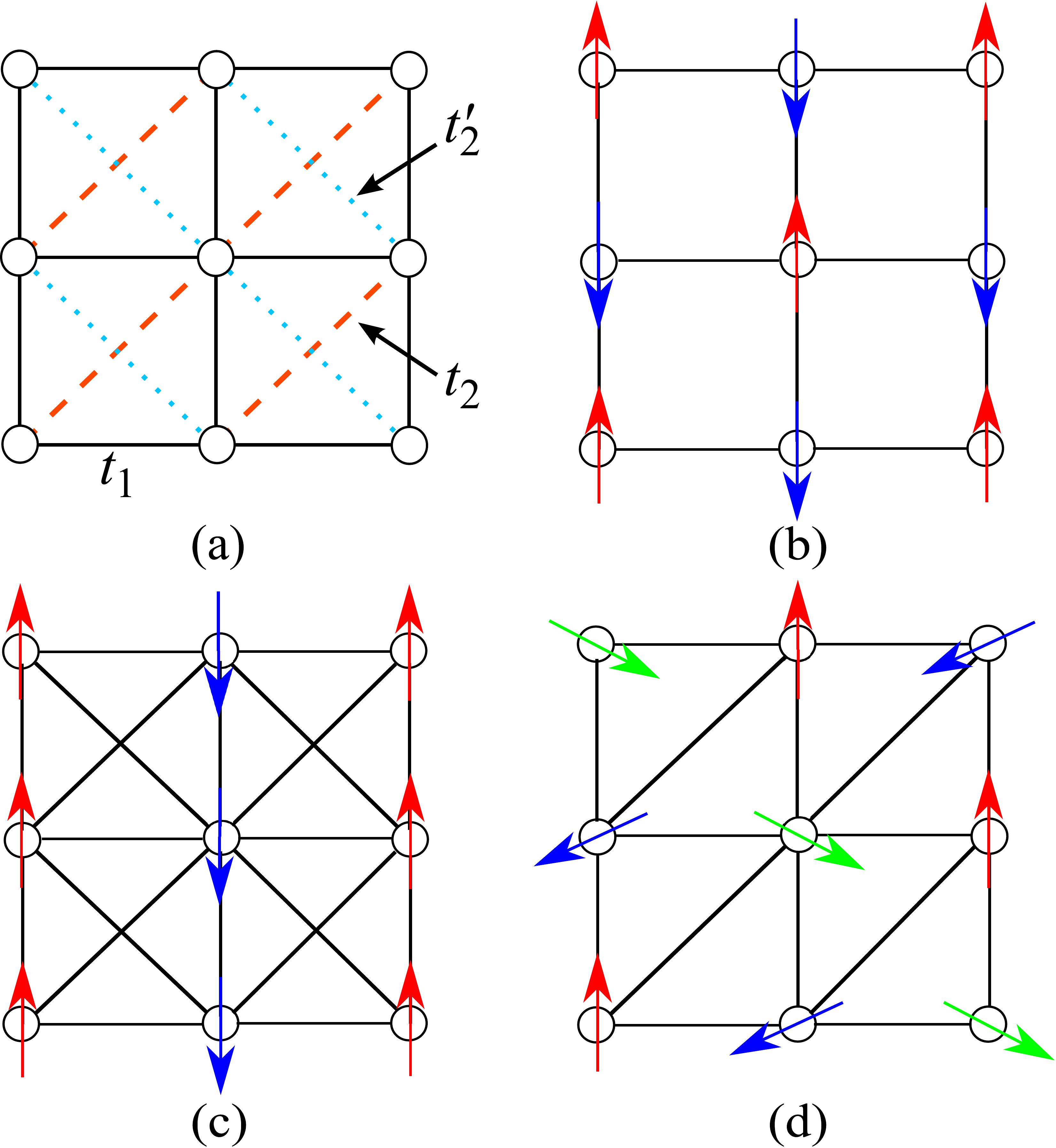}
\caption{(Color online) 
(a) Schematic representation of the square-lattice Hubbard 
model with the isotropic nearest-neighbor hopping parameter $t_1$ 
and anisotropic next-nearest-neighbor parameters $t_2$ and $t^{\prime}_2$.  
(b) Isotropic square lattice at $t_2=t^{\prime}_2=0$.  
(c) Crossed square lattice at $t_2=t^{\prime}_2=t_1$.  
(d) Isotropic triangular lattice at $t_1=t_2$ and $t^{\prime}_2=0$. 
The arrows represent the directions of the electron spins.  
The sublattices are indicated by different colors.  
}\label{fig1}
\end{center}
\end{figure}

\section{Model and Method}

We consider the frustrated Hubbard model defined on the two-dimensional 
square lattice at half filling as illustrated in Fig.~\ref{fig1}.  
The Hamiltonian is given by 
\begin{align}
&H=-t_1\sum_{ \langle i,j\rangle }\sum_{\sigma} c^{\dagger}_{i\sigma}c_{j\sigma} 
-t_2\sum_{\langle\langle i,j\rangle\rangle}\sum_{\sigma} c^{\dagger}_{i\sigma}c_{j\sigma} \notag \\ 
&-t^{\prime}_2\sum_{\langle\langle i,j\rangle\rangle^{\prime}}\sum_{\sigma} 
c^{\dagger}_{i\sigma}c_{j\sigma}
+U\sum_{i} n_{i\uparrow}n_{i\downarrow}
-\mu\sum_{i,\sigma} n_{i\sigma},
\label{ham}
\end{align} 
where $c^{\dagger}_{i\sigma}$ is the creation operator of an electron with 
spin $\sigma$ at site $i$ and $n_{i\sigma}=c^{\dagger}_{i\sigma}c_{i\sigma}$.  
$\langle i,j\rangle$ indicates the nearest-neighbor bonds with an isotropic 
hopping parameter $t_1$, and $\langle\langle i,j\rangle\rangle$ and 
$\langle\langle i,j\rangle\rangle^{\prime}$ indicate the next-nearest-neighbor 
bonds with anisotropic hopping parameters $t_2$ and $t^{\prime}_2$, respectively 
[see Fig.~\ref{fig1}(a)].  $U$ is the on-site Coulomb repulsion between 
electrons and $\mu$ is the chemical potential maintaining the system at half 
filling.  
In the large-$U$ limit, the model can be mapped onto the frustrated spin-1/2 
Heisenberg model 
\begin{align}
H=J_1\sum_{\langle i,j\rangle} \bm{S}_{i}\cdot \bm{S}_{j} 
+J_2\sum_{\langle\langle i,j\rangle\rangle} \bm{S}_{i}\cdot \bm{S}_{j}
+J^{\prime}_2\sum_{\langle\langle i,j\rangle\rangle^{\prime}} 
\bm{S}_{i}\cdot \bm{S}_{j} 
\label{hei}
\end{align}
in the second-order perturbation of the hopping parameters with 
$\bm{S}_i=\sum_{\alpha,\beta}c^\dagger_{i\alpha}\bm{\sigma}_{\alpha\beta}c_{i\beta}$/2, 
where $\bm{\sigma}_{\alpha\beta}$ is the vector of Pauli matrices.  
The exchange coupling constants are given by $J_1=4t^2_1/U$, $J_2=4t^2_2/U$, 
and $J^{\prime}_2=4t^{\prime 2}_2/U$ for the lattice shown in Fig.~\ref{fig1}(a).  
We will compare our results of the Hubbard model in the strong correlation 
regime with those of the frustrated Heisenberg model, for which related studies 
have been carried out.  

We treat a wide parameter space of $0\leq t_2/t_1\leq 1$ and 
$0\leq t^{\prime}_2/t_1\leq 1$, including three limiting cases: 
(i) at $t_2=t^{\prime}_2=0$ [square lattice, see Fig.~\ref{fig1}(b)], 
where the N\'eel order is realized, 
(ii) at $t_2=t^{\prime}_2=t_1$ [crossed square lattice, see Fig.~\ref{fig1}(c)], 
where the collinear order is realized, and 
(iii) at $t_2=t_1$ and $t^{\prime}_2=0$ [triangular lattice, 
see Fig.~\ref{fig1}(d)], where the 120$^{\circ}$ spiral order is realized.  
We will calculate how the above three ordered phases change 
when the hopping parameters are varied in the ranges $0\leq t_2\leq t_1$ and 
$0\leq t^{\prime}_2\leq t_1$.  

\begin{figure}[thb]
\begin{center}
\includegraphics[width=0.70\columnwidth]{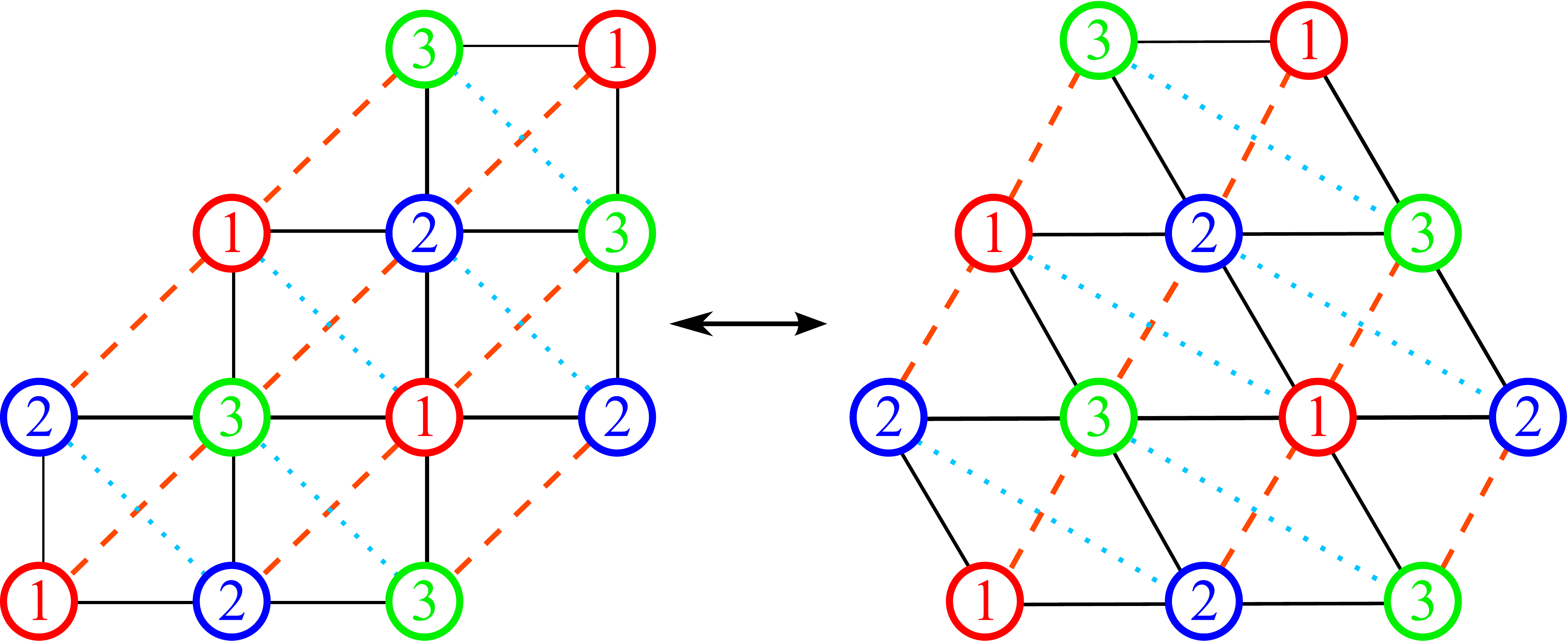}
\caption{(Color online) 
Left: twelve-site square-lattice cluster used as a reference system 
in our analysis.  Right: equivalent triangular-lattice cluster, 
where the three sites 1, 2, and 3 form an equilateral triangle.  
The anisotropic triangular lattice is defined as $t^{\prime}_2=0$ 
and $t_1\neq t_2$.  
}\label{fig2}
\end{center}
\end{figure}

We employ the VCA, which is a quantum cluster method based on SFT 
\cite{potthoff1,potthoff2,dahnken}, where the grand potential $\Omega$ 
of the original system is given by a functional of the self-energy.  
By restricting the trial self-energy to that of the reference system 
$\Sigma^{\prime}$, we obtain the grand potential in the thermodynamic 
limit as 
\begin{align}
\Omega[\Sigma']=\Omega'+\mathrm{Trln}(G_0^{-1}-\Sigma')^{-1}
-\mathrm{Trln}G^{\prime}, 
\label{self}
\end{align}
where $\Omega^{\prime}$ and $G^{\prime}$ are the exact grand potential 
and Green function of the reference system, respectively, and $G_0$ is 
the noninteracting Green function.  
The short-range electron correlations within the cluster of the reference 
system are taken into account exactly.  

The advantage of the VCA is that the spontaneous symmetry breaking can be 
treated within the framework of the theory.  Here, we introduce the Weiss 
fields for magnetic orderings as variational parameters.  The Hamiltonian 
of the reference system is then given by 
$H^{\prime}=H+H_{\rm{N}}+H_{\rm{C}}+H_{\rm{S}}$ with 
\begin{align}
&H_{\mathrm{N}}=h'_{\mathrm{N}}\sum_{i} e^{i\bm{Q}_{\mathrm{N}}\cdot \bm{r}_i} S^z_i \\
&H_{\mathrm{C}}=h'_{\mathrm{C}}\sum_{i} e^{i\bm{Q}_{\mathrm{C}}\cdot \bm{r}_i} S^z_i \\
&H_{\mathrm{S}}=h'_{\mathrm{S}}\sum_{i} \bm{e}_{a_i}\cdot \bm{S}_i ,
\label{vt}
\end{align}
where $h'_{\mathrm{N}}$, $h'_{\mathrm{C}}$, and $h'_{\mathrm{S}}$ are 
the strengths of the Weiss fields for the N\'eel, collinear, and spiral 
orders, respectively.  
The wave vectors are defined as $\bm{Q}_{\rm{N}}=(\pi,\pi)$ for the N\'eel 
order and $\bm{Q}_{\rm{C}}=(\pi,0)$ or $(0,\pi)$ for the collinear order.  
For the spiral order, the unit vectors $\bm{e}_{a_i}$ are rotated by 
120$^\circ$ to each other, where $a_i$ $(=1,2,3)$ is the sublattice index 
of site $i$.  The variational parameter is optimized on the basis of the 
variational principle $\partial\Omega/\partial h'=0$ for each magnetic 
order.  The solution with $h' \ne 0$ corresponds to the ordered state.  

We use the twelve-site cluster shown in Fig.~\ref{fig2} as the reference system.  
This cluster is convenient because we can treat the two-sublattice states 
(N\'eel and collinear states) with an equal number of up and down spins 
and, at the same time, the three-sublattice state (spiral state) with an 
equal number of three sublattice sites.  
Note that longer-period phases such as a spiral phase mentioned in a different 
system \cite{tocchio2} and incommensurate ordered phases are difficult to treat 
in the present approach.  

\begin{figure}[thb]
\begin{center}
\includegraphics[width=0.65\columnwidth]{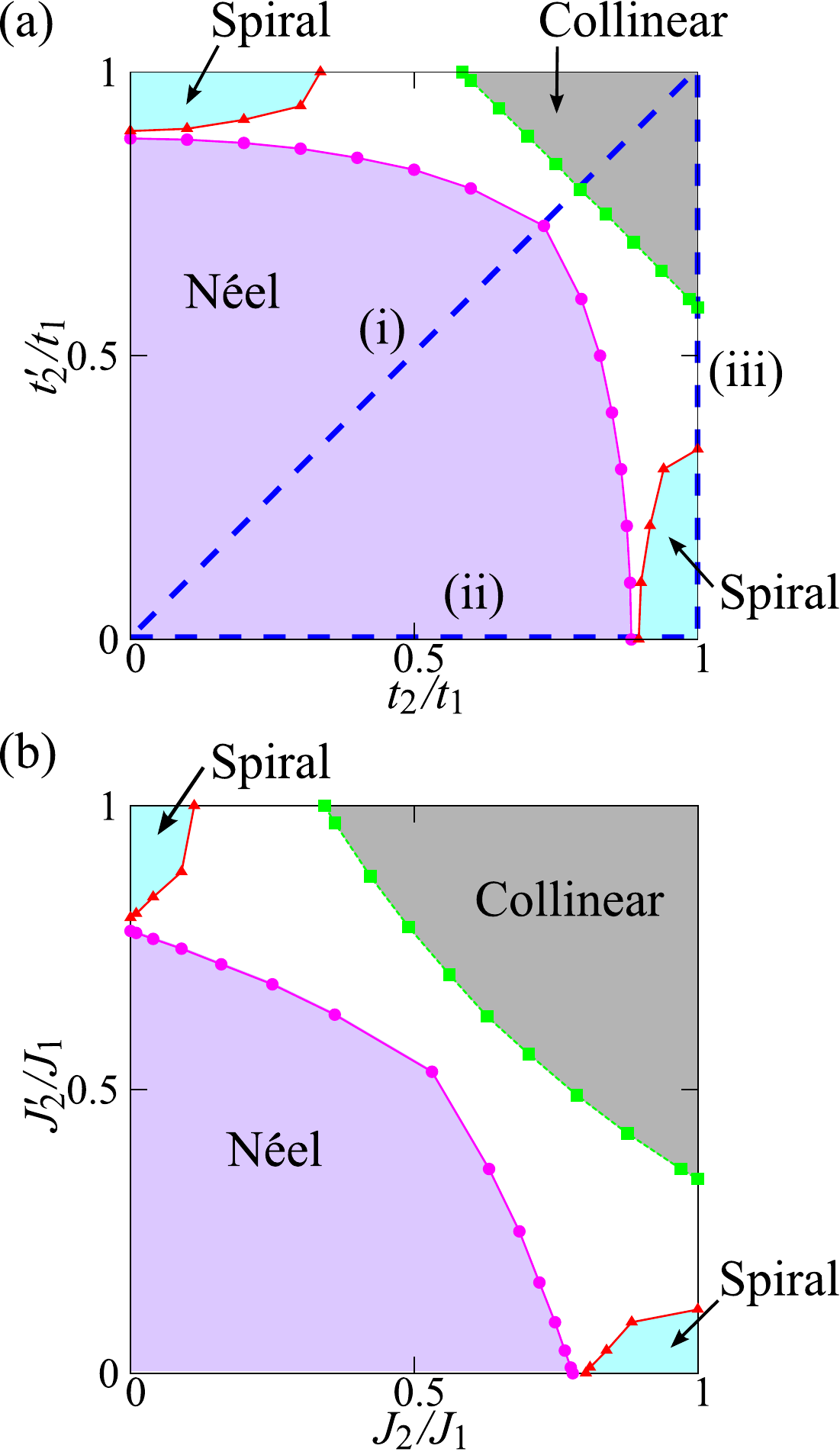}
\caption{(Color online) 
(a) Calculated ground-state phase diagram of our model at $U/t_1=60$ in the 
$(t_2/t_1, t^{\prime}_2/t_1)$ plane and (b) converted phase diagram in the 
$(J_2/J_1, J^{\prime}_2/J_1)$ plane.  
The uncolored region corresponds to the nonmagnetic insulating phase.  
The transition to the collinear phase is of the first order (or discontinuous) and 
the transitions to the N\'eel and spiral phases are of the second order (or continuous).  
The phases along dashed lines (i), (ii), and (iii) shown 
in (a) are circumstantiated in Fig.~\ref{fig4}.  
}\label{fig3}
\end{center}
\end{figure}

\begin{figure*}[thb]
\begin{center}
\includegraphics[width=2.0\columnwidth]{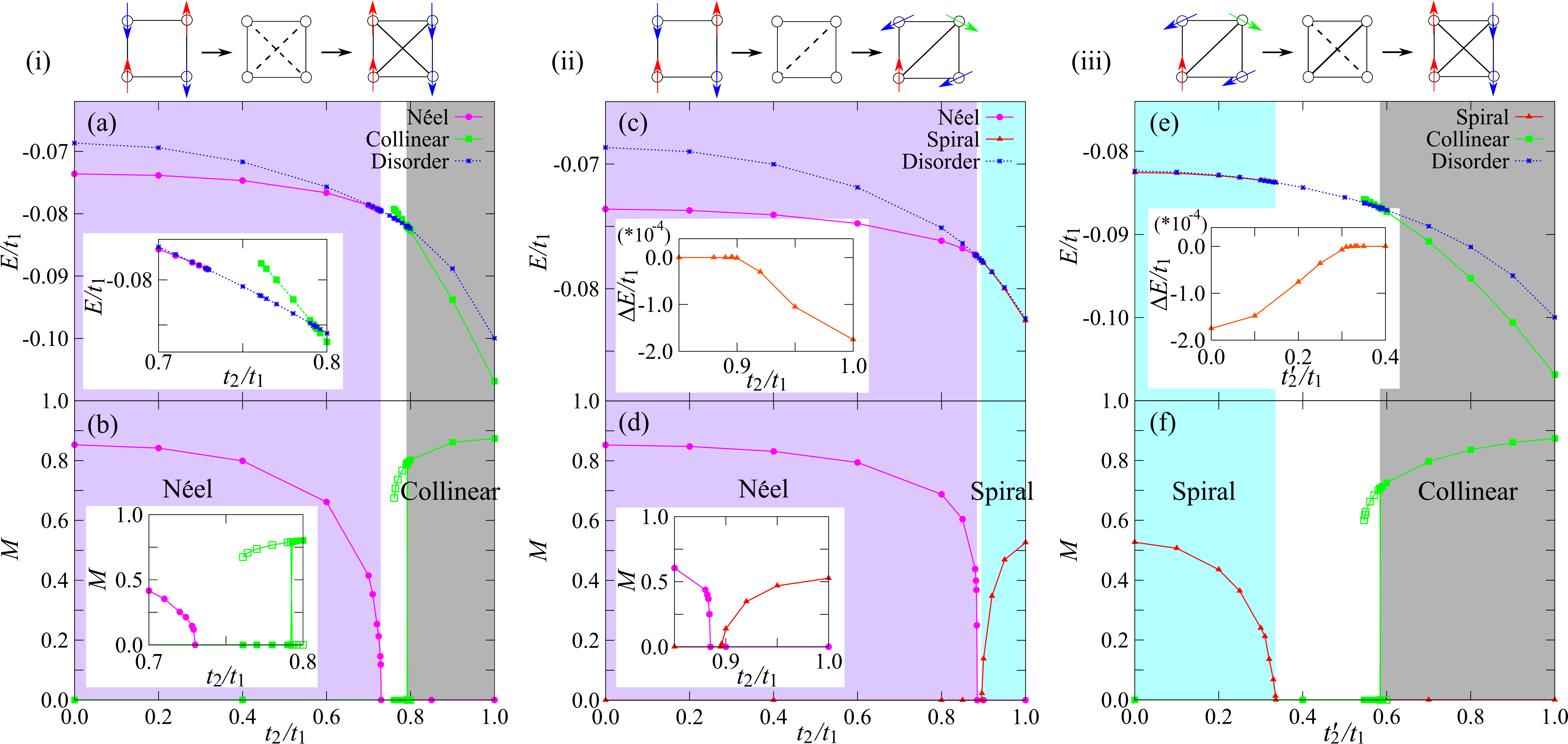}
\caption{(Color online) 
Calculated ground-state energies (upper panels) and order parameters (lower panels) 
for the N\'eel, collinear, spiral, and nonmagnetic insulating phases as a function of 
$t_2/t_1$ or $t^{\prime}_2/t_1$.  The left, middle, and right panels correspond to lines 
(i), (ii), and (iii) in Fig.~\ref{fig3}(a), where we assume $t_2=t^{\prime}_2$, 
$t^{\prime}_2=0$, and $t_2=t_1$, respectively.  The inset in (c) and (e) displays the energy 
difference between the spiral and nonmagnetic insulating phases $\Delta E$, and other 
insets enlarge the region near the phase boundary.  
}\label{fig4}
\end{center}
\end{figure*}

\section{Results of Calculations}

\subsection{Strong correlation regime}

First, let us discuss the phase diagram of our model in the strong correlation regime 
$U/t_1=60$.  The result is shown in Fig.~\ref{fig3}, where the result for our Hubbard 
model in the $(t_2/t_1, t^{\prime}_2/t_1)$ plane as well as the same result converted 
to the Heisenberg model parameters $(J_2/J_1, J^{\prime}_2/J_1)$ are shown.  
We find three ordered phases:  
the N\'eel ordered phase around $(t_2/t_1, t^{\prime}_2/t_1)=(0,0)$, 
the collinear ordered phase around $(t_2/t_1, t^{\prime}_2/t_1)=(1,1)$, and 
the spiral ordered phase around $(t_2/t_1, t^{\prime}_2/t_1)=(1,0)$ and 
$(0,1)$.  
The nonmagnetic insulating phase, which is absent in the classical Heisenberg model (see Appendix A), 
appears in an intermediate region between the three ordered phases.  
Thus, the quantum fluctuations in the frustrated spin degrees of freedom are essential 
in the emergence of the nonmagnetic insulating phase.  
As shown below, the phase transition to the collinear phase is of 
the first order (or discontinuous) and the phase transitions to 
the N\'eel and spiral phases are of the second order (or continuous).  
This phase diagram is determined on the basis of the calculated ground-state 
energies $E=\Omega+\mu$ (per site) and magnetic order parameters $M$ 
(per site) defined as 
$M_{\rm{N}}=(2/L)\sum_{i} e^{i\bm{Q}_{\mathrm{N}}\cdot \bm{r}_i} \langle S^z_i\rangle$ 
for the N\'eel order, 
$M_{\rm{C}}=(2/L)\sum_{i} e^{i\bm{Q}_{\mathrm{C}}\cdot \bm{r}_i} \langle S^z_i \rangle$ 
for the collinear order, and 
$M_{\rm{S}}=(2/L)\sum_{i} \bm{e}_{a_i}\cdot \langle \bm{S}_i \rangle$ 
for the spiral order, 
where $\langle~\rangle$ stands for the ground-state expectation value and $L$ is 
the number of sites in the system.  
In the following, we will circumstantiate the obtained phases, particularly along 
lines (i), (ii), and (iii) drawn in Fig.~\ref{fig3}(a), whereby 
we will discuss some details of our calculated results in comparison with other studies.  

Along line (i):  
The results are shown in the left panels of Fig.~\ref{fig4}, where 
we assume $t_2=t^{\prime}_2$.  At $t_2=0$, the ground state is 
the N\'eel order, and with increasing $t_2$, the energy of the N\'eel 
order gradually approaches the energy of the nonmagnetic state.  
At $t_2/t_1=0.73$, the energy of the N\'eel order continuously 
reaches the energy of the nonmagnetic state and the N\'eel order 
disappears.  The calculated order parameter indicates a continuous phase transition.  
At $t_2/t_1=1$, on the other hand, the ground state is the collinear order.  
The ground-state energy of the collinear order increases with decreasing $t_2$, and 
at $t_2/t_1=0.79$, it crosses to the nonmagnetic state, resulting 
in a discontinuous phase transition, as indicated by the calculated order parameter.  
The nonmagnetic insulating state thus appears at $0.73 < t_2/t_1 < 0.79$, 
which corresponds to the region $0.53 < J_2/J_1 < 0.63$ in the 
Heisenberg model parameters.  
In comparison with previous studies on the $J_1$-$J_2$ square-lattice 
Heisenberg model, which have estimated the transition point between 
the N\'eel and nonmagnetic phases to be at $J_2/J_1=0.40-0.44$ 
\cite{darradi,yu1,jiang2,gong,morita1}, our result slightly 
overestimates the stability of the N\'eel order.  
This overestimation may be caused by the cluster geometry used 
in our calculations; if we use the $2\times 2$ site cluster as 
the reference system, the transition occurs at $J_2/J_1 = 0.42$ 
\cite{yamaki}, which is in good agreement with the previous 
studies.  
The transition point between the collinear and nonmagnetic phases, 
on the other hand, has been estimated to be at $J_2/J_1=0.59-0.62$ 
\cite{darradi,yu1,jiang2,morita1}, which is in good agreement 
with our result.  

Along line (ii):  
The results are shown in the middle panels of Fig.~\ref{fig4}, 
where we assume $t^{\prime}_2=0$.  
With increasing $t_2$ from $t_2=0$, at which the ground state 
is the N\'eel order, the energy of the N\'eel order gradually 
approaches the energy of the nonmagnetic state, and at $t_2/t_1=0.88$, 
the N\'eel order disappears continuously.  The calculated order 
parameter indicates the continuous phase transition. 
At $t_2/t_1=1$, on the other hand, the ground state is the spiral 
order, although the energy difference between the spiral and 
nonmagnetic states is very small [see the inset of Fig.~\ref{fig4}(c)] 
due to the strong geometrical frustration of the triangular 
lattice.  
With decreasing $t_2$ from $t_2/t_1=1$, the ground-state energy 
of the spiral order increases gradually and approaches the energy 
of the nonmagnetic state, and at $t_2/t_1 = 0.89$, the spiral order 
disappears continuously, in agreement with the calculated order 
parameter.  Thus, the nonmagnetic phase appears in a very narrow 
region of $0.88 < t_2/t_1 < 0.89$.  The corresponding Heisenberg model 
parameters at which the N\'eel and spiral orders disappear are 
around $J_2/J_1 = 0.79$.  
The previous studies for the anisotropic triangular-lattice Heisenberg 
model \cite{weng,bishop} have given values around $J_2/J_1 = 0.80 - 0.87$ 
for the transition point, which are in good agreement with our result.  

\begin{figure*}[bht]
\begin{center}
\includegraphics[width=1.65\columnwidth]{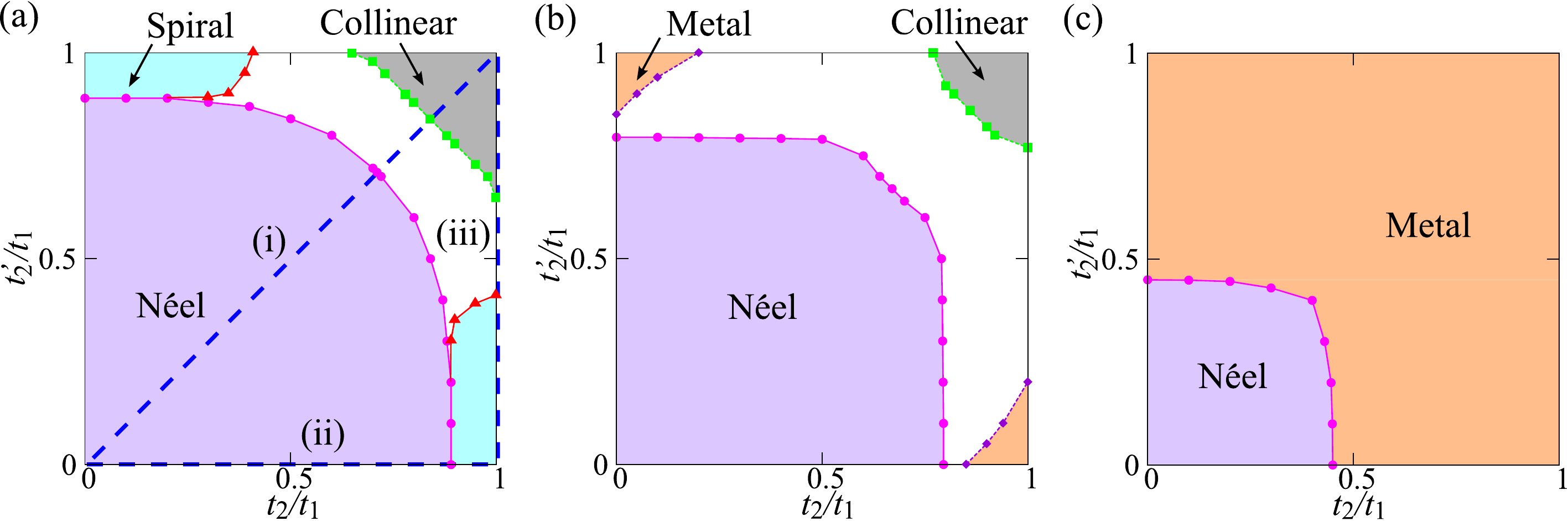}
\caption{(Color online) 
Calculated ground-state phase diagrams of our model in the $(t_2/t_1, t^{\prime}_2/t_1)$ plane at 
(a) $U/t_1=10$, (b) $U/t_1=6$, and (c) $U/t_1=2$.  The uncolored regions in (a) and (b) correspond 
to the nonmagnetic insulating phase.  The phases along dashed lines (i), (ii), and (iii) 
shown in (a) are circumstantiated in Fig.~\ref{fig6}.  
}\label{fig5}
\end{center}
\end{figure*}

\begin{figure*}[thb]
\begin{center}
\includegraphics[width=2.0\columnwidth]{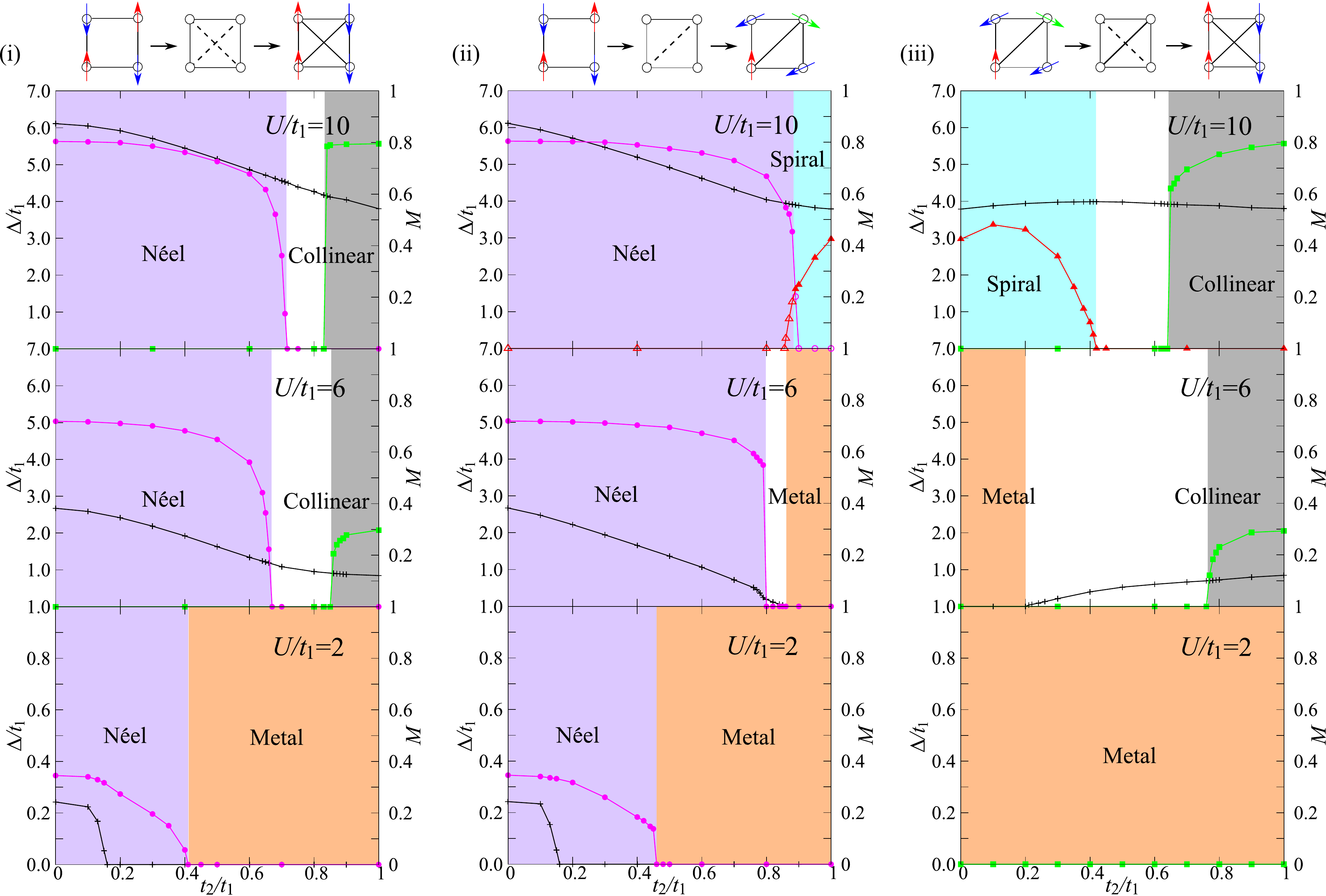}
\caption{(Color online) 
Calculated results for the order parameters $M$ (pink, green, and red dots) of the N\'eel, collinear, 
and spiral phases and the single-particle gap $\Delta/t_1$ (black crosses) at $U/t_1=10$ (upper panels), 
$U/t_1=6$ (middle panels), and $U/t_1=2$ (lower panels).  The left, middle, and right panels correspond 
to the lines (i), (ii), and (iii) defined in Fig.~\ref{fig5}(a), where we assume $t_2=t^{\prime}_2$, 
$t^{\prime}_2=0$, and $t_2=t_1$, respectively. 
}\label{fig6}
\end{center}
\end{figure*}

Along line (iii):  
The results are shown in the right panels of Fig.~\ref{fig4}, where 
we assume $t_2=t_1$.  At $t^{\prime}_2 = 0$, the ground state is 
the spiral order, although the energy difference from the nonmagnetic 
state is very small [see the inset of Fig.~\ref{fig4}(e)].  
With increasing $t^{\prime}_2$, the energy of the spiral order 
gradually approaches the energy of the nonmagnetic state, and 
at $t^{\prime}_2/t_1 = 0.34$, the spiral order disappears 
continuously, in agreement with the calculated order parameter.  
On the other hand, with decreasing $t^{\prime}_2$ from 
$t^{\prime}_2/t_1=1$, at which the collinear order is stable, 
the ground-state energy of the collinear order increases and 
crosses to the nonmagnetic state at $t^{\prime}_2/t_1=0.59$.  
The transition is thus discontinuous, in agreement with the 
calculated order parameter.  The nonmagnetic state therefore 
appears at $0.34 < t^{\prime}_2/t_1 < 0.59$, which corresponds 
to the region $0.11 < J^{\prime}_2/J_1 < 0.34$ if we use the 
Heisenberg model parameters.  To our knowledge, no comparable 
calculations have been made for the frustrated Heisenberg model 
in this parameter region.  

\subsection{Intermediate to weak correlation regime}

Next, let us discuss the phase diagram of our model in the intermediate to weak correlation regime.  
The results at $U/t_1=10$, $6$, and $2$ are shown in Fig.~\ref{fig5}.  The detailed results for the 
calculated single-particle gap and order parameters are also shown in Fig.~\ref{fig6} along lines 
(i), (ii), and (iii) defined above.  

At $U/t_1=10$, we find that the results are qualitatively similar to those at $U/t_1=60$, except 
for the transition between the N\'eel and spiral orders: the nonmagnetic insulating phase appears 
between these orders at $U/t_1=60$ but a direct first-order transition occurs at $U/t_1=10$ 
\cite{yamada,laubach} with a double minimum structure in the grand potential.  
The nonmagnetic insulating phase appears at $0.71 < t_2/t_1 < 0.84$ along line (i), which is in 
good agreement with the previous studies \cite{nevidomskyy,yamaki}, where the values 
$t_2/t_1=0.70-0.77$ for the transition between the N\'eel and nonmagnetic phases and 
$t_2/t_1=0.82-0.85$ for the transition between the collinear and nonmagnetic phases were reported.  
The nonmagnetic phase also appears at $0.42 < t^{\prime}_2/t_1 < 0.64$ along line (iii).  

At $U/t_1=6$, we find that the spiral phase disappears and a paramagnetic metallic phase 
appears in the triangular lattice geometry at $(t_2/t_1, t_2'/t_1)\simeq(1,0)$ or $(0,1)$.  
The nonmagnetic insulating phase appears at $0.66 < t_2/t_1 < 0.86$ along line (i), the region of 
which becomes wider with decreasing value of $U/t_1$ from 60 to 10 and 6, which is again in 
good agreement with the previous studies \cite{mizusaki,nevidomskyy}.  
The region of the nonmagnetic insulating phase also becomes wider along line (iii), which 
occurs between the collinear and paramagnetic metallic phases.  
Along line (ii), the nonmagnetic insulating phase with a small charge gap appears again, which 
is between the N\'eel and paramagnetic metallic phases \cite{yamada, laubach}.  

At $U/t_1=2$, the paramagnetic metallic phase overwhelms the collinear and nonmagnetic insulating 
phases, retaining only the N\'eel ordered phase around $t_2/t_1=t_2'/t_1=0$.  Within the N\'eel 
phase, the charge gap opens only at $0<t_2/t_1<0.16$ and the metallic N\'eel ordered phase 
appears at $0.16<t_2/t_1<0.41$ along line (i) and at $0.16<t_2/t_1<0.47$ along line (ii).  
The perfect Fermi surface nesting at $t_2/t_1=t_2'/t_1=0$ and its deformation away from 
$t_2/t_1=t_2'/t_1=0$ are responsible for these results \cite{nevidomskyy}.  
The generalized magnetic susceptibility $\chi_0(\bm{q})$ calculated in the noninteracting limit of 
our model [Eq.~(\ref{ham})] explains this result (see Appendix B).   
The transition between the N\'eel ordered metallic phase and the paramagnetic metallic phase 
is continuous along line (i) and discontinuous along line (ii).  

\bigskip
\section{Summary}

We have used the VCA based on SFT to study the two-dimensional frustrated Hubbard 
model at half filling with the isotropic nearest-neighbor and anisotropic next-nearest-neighbor 
hopping parameters.  We have particularly focused on the effect of geometrical frustration 
on the spin degrees of freedom of the model in a wide parameter space including square, 
crossed-square, and triangular lattices in a wide range of the interaction strength at zero 
temperature.  We have thereby investigated the metal-insulator transition, the magnetic orders, 
and the emergence of the nonmagnetic insulating phase, although the phases with incommensurate 
orders or with longer-period orders than the cluster size used have not been taken into account 
owing to the limitation of the VCA.  
We have also calculated the ground-state phase diagram of the corresponding classical 
Heisenberg model as well as the generalized magnetic susceptibility in the noninteracting 
limit.  

We have thus determined the ground-state phase diagram of the model and found that, in the 
strong correlation regime, magnetic phases with the N\'eel, collinear, and spiral orders 
appear in the parameter space, and a nonmagnetic insulating phase, caused by the effect of 
quantum fluctuations in the frustrated spin degrees of freedom, emerges in the wide parameter 
region between these three ordered phases.  
We have also found that the phase transition from the N\'eel and spiral orders to the 
nonmagnetic phase is continuous (or a second-order transition), whereas the transition 
from the collinear order to the nonmagnetic phase is discontinuous (or a first-order transition).  
We have compared our results with the results of the corresponding Heisenberg model 
calculations that have been made so far and found that the agreement is good whenever the 
comparison is possible.  
We have also found that, in the intermediate correlation regime, the paramagnetic metallic 
phase begins to appear in the triangular lattice geometry, which overwhelms the collinear 
and nonmagnetic insulating phases in the weak correlation regime, retaining only the 
N\'eel ordered phase in the square lattice geometry.  
We hope that our results for the phase diagram obtained in the wide parameter space will 
encourage future studies on the characterization of the nonmagnetic insulating phase as well as 
on its experimental relevance.  

\begin{acknowledgment}
We thank S.~Miyakoshi for useful discussions.  T.~K. acknowledges 
support from a JSPS Research Fellowship for Young Scientists.  
This work was supported in part by a Grant-in-Aid for Scientific 
Research (No.~26400349) from JSPS of Japan. 
\end{acknowledgment}

\appendix

\section{Ground-State Phase Diagram of the Classical Heisenberg Model}

\begin{figure}[thb]
\begin{center}
\includegraphics[width=0.62\columnwidth]{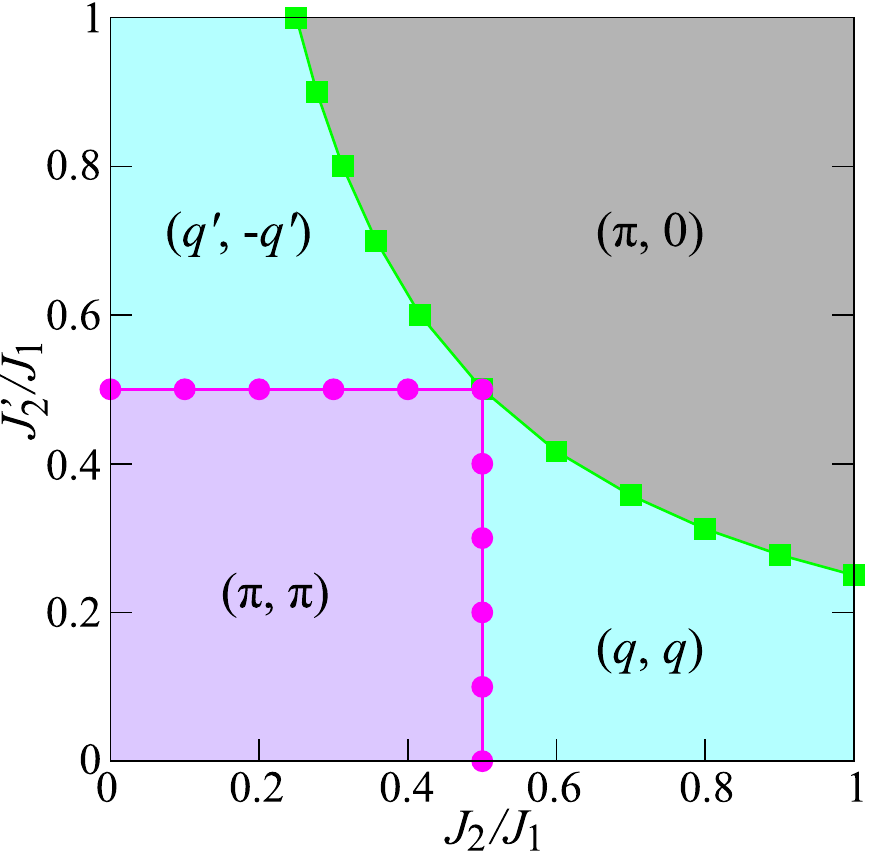}
\caption{(Color online) 
Calculated ground-state phase diagram of the corresponding classical Heisenberg model.  
}\label{fig7}
\end{center}
\end{figure}

Here, we present the ground-state phase diagram of the classical Heisenberg model, which is 
defined as in Eq.~(2) but its quantum spins $\bm{S}_i$ are replaced by the classical vectors 
$\tilde{\bm{S}}$, so that quantum fluctuations of the system are completely suppressed 
although the frustrative features in the spin degrees of freedom are present.  The Hamiltonian 
is given by $H=\sum_{\bm{q}}J(\bm{q})\,\tilde{\bm{S}}_{-\bm{q}}\cdot\tilde{\bm{S}}_{\bm{q}}$ 
in momentum space, where 
\begin{align}
J(\bm{q})&=J_1(\cos q_x+\cos q_y)+J_2\cos(q_x+q_y) \notag \\
&+J^{\prime}_2\cos(q_x-q_y). 
\label{ssf}
\end{align}
The ground states of the system are calculated \cite{luttinger} and the phase diagram is 
obtained as shown in Fig.~\ref{fig7}.  We find that the magnetically ordered ground states appear 
in the entire parameter space examined, which include 
the N\'eel order [$\bm{q}=(\pi,\pi)$], 
collinear order [$\bm{q}=(\pi,0)$], and 
spiral orders [$\bm{q}=(q,q)$ and $(q^{\prime},-q^{\prime})$ with $q=\cos^{-1}(-J_1/2J_2)$ and 
$q^{\prime}=\cos^{-1}(-J_1/2J^{\prime}_2)$].  
Therefore, comparing with the results given in the main text, we may conclude that the 
quantum fluctuations in the geometrically frustrated spin degrees of freedom are 
essential in the emergence of the nonmagnetic insulating phase discussed in the main text.  

\section{Generalized Susceptibility in the Noninteracting Limit}

\begin{figure}[thb]
\begin{center}
\includegraphics[width=1.0\columnwidth]{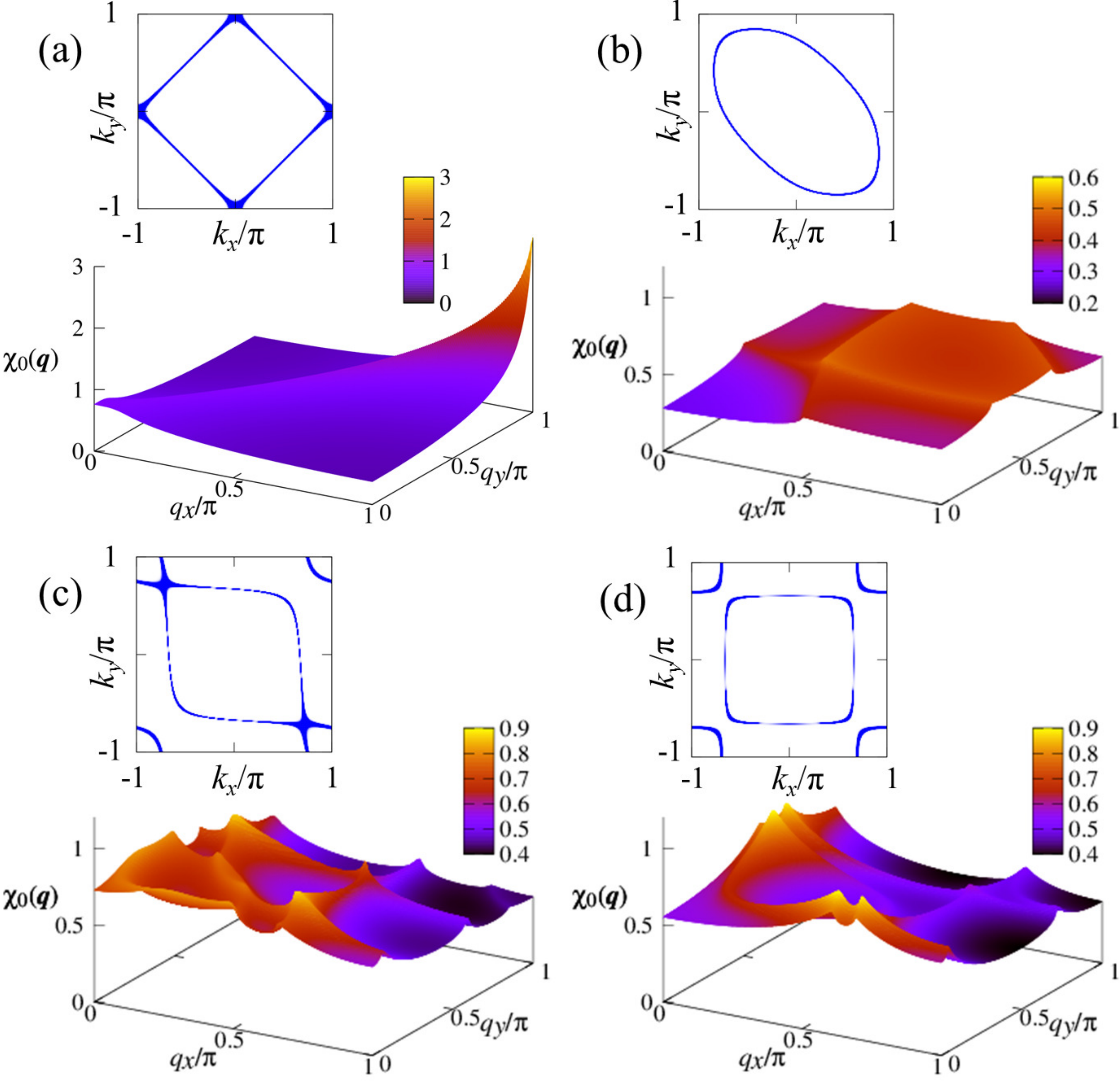}
\caption{(Color online) 
Calculated generalized magnetic susceptibility defined in Eq.~(\ref{chiq}) at 
(a) $t_2/t_1=t^{\prime}_2/t_1=0.0$, 
(b) $t_2/t_1=1.0$, $t^{\prime}_2/t_1=0.0$, 
(c) $t_2/t_1=1.0$, $t^{\prime}_2/t_1=0.8$, and 
(d) $t_2/t_1=t^{\prime}_2/t_1=1.0$.  
The corresponding Fermi surface is shown in each panel.  
}\label{fig8}
\end{center}
\end{figure}

Here, we present the generalized magnetic susceptibility (or Lindhard function) at zero frequency 
\cite{hirsch,bulut,sherman}, 
\begin{align}
\chi_0(\bm{q})=\frac{1}{L}\sum_{\bm{k}} \frac{f(\epsilon_{\bm{k}})-f(\epsilon_{\bm{k}+\bm{q}})}
{\epsilon_{\bm{k}+\bm{q}}-\epsilon_{\bm{k}}} , 
\label{chiq}
\end{align}
calculated for our model [Eq.~(1)] in the noninteracting limit, where $\epsilon_{\bm{k}}$ is the 
corresponding noninteracting band disperson and $f(\epsilon)$ is the Fermi function.  
The calculated results at temperature $0.01t_1$ are shown in Fig.~\ref{fig8}, where we find 
that a diverging behavior appears only at $\bm{q}=(\pi,\pi)$ in Fig.~\ref{fig8}(a) due to perfect 
Fermi surface nesting, which yields the N\'eel ordered state at $t_2/t_1=t^{\prime}_2/t_1=0.0$ in 
the presence of a small but finite interaction strength $U$.  There are characteristic features 
of $\chi_0(\bm{q})$ but no other diverging behaviors are found, indicating the absence of other 
magnetic orderings in the weak correlation limit.

\end{document}